\documentclass[12pt]{article}
\usepackage{graphicx}

\title{NONEQUILIBRIUM BOUND INTERFACES}
\author{F. de los Santos$^1$ and M.M. Telo da Gama$^2$}

\date{$^1$Departamento de Electromagnetismo y F\'\i sica de la
Materia, Universidad de Granada, Fuentenueva s/n, 18071 Granada,
Spain \\ $^2$Centro de F\'\i sica Te\'orica e Computacional e\\ 
Departamento de F\'\i sica da Faculdade de Ci\^encias da Universidade de 
Lisboa, Av. Prof. Gama Pinto, 2, P-1643-003 Lisboa
Codex, Portugal}

\begin{document}
\maketitle
\pagebreak

\begin{abstract}

An overview of recent studies of nonequilibrium 
bound interfaces is given. Attention is focused 
on Kardar-Parisi-Zhang interfaces in the presence of 
upper and lower walls, interacting via short-- and 
long--ranged potentials. 
A comparison with equilibrium interfaces is carried out, and
connections with other nonequilbrium systems are illustrated.
Experimental realizations of the phenomenology described in this 
article are briefly discussed.

\end{abstract}

\section{INTRODUCTION}
An interface is the moving or static boundary between two distinct
bulk phases. Effective descriptions that focus on the statistical 
properties of the interfacial degrees of freedom have 
proved particularly useful, and over the past decades  
received a great deal of attention \cite{barabasi,meakin,jasnow,hz}.
Within this approach, the essence of the bulk contributions is 
incorporated into the (continuous or discrete) model parameters; other 
effects, such as geometric constraints, are modeled as nontrivial 
surface terms. Examples of the latter abound, and include 
the morphology and motion of an interface constrained by the 
presence of a third phase, by an external 
field, by inhomogeneities of the medium, etc. In thermal equilibrium,
or near equilibrium, enormous theoretical, as well as experimental, 
progress was made. For example, 
wetting in the two-dimensional Ising model was shown  
to be uniquely described by an interfacial model on sufficiently large 
length scales \cite{upton}. Of more practical interest 
are three-dimensional systems with realistic interactions, for which  
a number of theoretical predictions have been confirmed experimentally
\cite{reviewbonn}. However, for nonequilibrium systems this field of 
research is only just starting. 

In this review, we give an overview of recent theoretical results on
nonequilibrium bound interfaces. The focus will be on a few
models, mostly in the form of continuum stochastic growth
equations, that are expected to be representative of a broad class of
physical phenomena. The primary reason for this choice is that,
the statistical properties being identical to those of the discrete
models, continuum equations are more adequate to describe the collective 
behavior at the macroscopic level, and thus give a unified 
picture of the field.
Although the discussion is intended to be general, unless 
otherwise stated, results pertaining to one-dimensional systems only are 
given.

This paper is organized as follows. Firstly, in Section 2 a brief 
discussion of the kinetics of two representative, free-interface models 
is provided. In Section 3, we review the implications 
brought about by the inclusion of interacting, binding walls.  
Next, we explore analogous phenomena in other systems and in the last 
section describe experimental systems where bound-KPZ behavior could  
be observed. 
Much of this material has been covered from a different perspective 
in a recent review \cite{review_munoz}.

\section{TWO MODELS OF FREE INTERFACES}
\subsection{The Edwards-Wilkinson equation}

The following Langevin equation, referred to as the
Edwards-Wilkinson equation, is commonly employed to describe on a
coarse-grained level the dynamics of an interface separating two
bulk fluid phases \cite{ew},

\begin{equation}
{\partial h({\bf x},t)\over \partial t} = \nu \nabla^2h  +
\eta({\bf x},t), \label{eweq}
\end{equation}
where $h({\bf x},t)$ is the local height of the interface above
some reference plane. $\nu$ is the interfacial tension or the interfacial 
stiffness if any of the coexisting phases is anisotropic. It measures 
the strength of the Laplacian, a relaxing term that arises from bulk 
evaporation-condensation processes. Microscopic
fluctuations, which in the present case are assumed to have a
thermal origin, are modeled by the stochastic noise term $\langle
\eta \rangle$, with mean $\langle \eta \rangle =0$ and variance
$\langle \eta({\bf x},t) \eta({\bf x}',t') \rangle =
2D\delta(t-t') \delta({\bf x}-{\bf x}')$, where $\langle \cdot
\rangle$ denotes an average over the noise distribution.
Since the system is in thermal equilibrium, the noise correlation
function is related to the temperature through the
fluctuation-dissipation theorem, i.e. $D=k_BT$. Finally, note that
this model ignores configurations containing overhangs and
bubbles, as $h({\bf x},t)$ is a single-valued function.

An interface governed by the EW equation does not move on average
irrespective of its initial position. This agrees with the
thermodynamic picture that at bulk coexistence any arbitrary
fraction of the system may be in one phase, with the remainder in
the other. Adding a constant $\mu$ to (\ref{eweq}) drives the
system out of coexistence. For instance, if $\mu>0$ the
steady-state interface moves upwards with an
average velocity $v=\mu$, thereby favoring the $y<h({\bf x})$
phase over the unstable $y>h({\bf x})$ one. In this context,
$\mu$ may be interpreted as an external force acting on the
interface or as the chemical potential difference between the two
phases.

The fluctuations of the interface around its equilibrium position
are characterized by the interfacial width, a
conventional measure of interfacial roughness, defined as
\begin{equation}
W^2(L,t) = \langle  [ h(t) - \bar{h}(t)]^2  \rangle,
\end{equation}
where $\bar{h}(t)$ is the spatial average of $h({\bf x},t)$. An
initially flat surface is essentially confined
between $\bar{h}(t) - W(t)$ and $\bar{h}(t)+W(t)$, at later times. It is
well-known that the behavior of $W$ depends on the system
dimensionality: for $d>2$ the interface is said to be smooth
as $W$ converges in the limit $L,t \to \infty$. At $d=2$ the interface is 
logarithmically rough,
\begin{equation}
W^2 \sim \cases { \ln(\nu t/ a), \qquad t\ll L^z, \cr \ln(L/ a),
\qquad t\gg L^z.}
\end{equation}
Here $a$ is a microscopic scale of the order of the bulk
correlation length and $L^z$ defines a time scale characterized
by the so-called {\em dynamic exponent} $z$. Finally, in $d<2$ the 
interfacial behavior is universal and given by,
\begin{equation}
W \sim \cases {t^\beta, \qquad t\ll L^z, \cr L^{\zeta}, \qquad
t\gg L^z.} 
\label{width}
\end{equation}
The initial increase of the interfacial width as a power of time is
governed by the {\em growth exponent}, $\beta$; this is followed
by a saturation regime where the interfacial width scales as $W(L,\infty) 
\sim L^\zeta$.
The crossover between the initial and the saturation regimes occurs  
when the lateral correlation length
$\xi(t) \sim t^{1/z}$ is of the order of the size of the system. 
The {\em roughness exponent} $\zeta$, is a measure of the typical
transverse fluctuation $W$ of a given interfacial segment $L$ as
given by (\ref{width}). This behavior is summarized by the
Family-Vicseck scaling relation $W(L,t) \sim L^\zeta f(t/L^z)$,
where the scaling function $f$ satisfies $f(u) \sim u^\beta$ for
$u \ll 1$ and $f(u)=const$ for $u \gg 1$ \cite{family-vicsek}.
Clearly, continuity at $t=L^z$ requires $\zeta= \beta z$. Note that
in an infinite system the width does not saturate, implying
that the interface is rough in the sense that it makes arbitrarily
large excursions from its average position. Due to the linear
character of equation (\ref{eweq}) the exponents can be
calculated in arbitrary dimension. In particular, $z=2,
\beta=(2-d)/4$, and $\zeta=(2-d)/2$.

Equation (\ref{eweq}) can be derived by differentiating a
simple Hamiltonian
\begin{equation}
{\cal H}(h)=\int_0^\infty d{\bf x}\bigg[ {\nu \over 2} (\nabla
h)^2 \bigg] \label{ham}
\end{equation}
that is recognized as the square-gradient approximation to a surface free 
energy that is proportional to the total interfacial area,
\begin{equation}
{\cal F} = \nu \int d{\bf x}\bigg[  1+(\nabla h)^2 \bigg]^{1/2}.
\end{equation}
The study of systems off-coexistence requires the inclusion of an 
additional term
$\mu
\int h({\bf x}) d{\bf x}$ in (\ref{ham}), where $\mu$ is the chemical 
potential difference between the two phases. Of
course, all the stationary properties of the EW interface described
above can be recovered from (\ref{ham}) in a straightforward manner.

\subsection{The Kardar-Parisi-Zhang equa\-tion}

As referred previously, the velocity of an interface
described by the EW equation is $v=\mu$, regardless of its initial
position. At $\mu =0$ this is equivalent to bulk phase coexistence. A 
natural generalization of
(\ref{eweq}) consists in assuming that $v$ is no longer constant,
but depends on the instantaneous local-slope of the interface at
${\bf x}$. Thus, substituting $\mu$ by $v(\nabla h)$ and using
again the small-slope approximation one finds\footnote{It can be
shown that third and higher order terms may be neglected in
the large-scale limit \cite{maritan}.}
\begin{equation}
v(\nabla h) \approx v({\bf 0}) +\nabla v({\bf 0}) \cdot \nabla h +
{1 \over 2} [(\nabla h)\cdot \nabla]^2 v(\nabla h=0).
\end{equation}
Assuming isotropy in the sense that $\partial^2 v /\partial
x_i\partial x_j \propto \delta_{ij}$, and removing the first two
terms through a Galilean transformation $h({\bf x},t) \to h({\bf
x}+\nabla v({\bf 0}) t,t) +v({\bf 0})t$, yields the celebrated
Kardar-Parisi-Zhang (KPZ) equation \cite{kpz},
\begin{equation}
{\partial h({\bf x},t)\over \partial t} = \nu \nabla^2 h  +
\lambda (\nabla h)^2 + \eta({\bf x},t). \label{kpzeq}
\end{equation}
The only difference between the KPZ and the EW equations is the
nonlinear term $\lambda (\nabla h)^2$, that is the most 
relevant nonequilibrium perturbation to the equilibrium EW equation and 
is expected to describe the scaling properties of rough surfaces growing 
in the absence of conservation laws.
Somewhat surprisingly, experimental realizations of the KPZ behavior are 
far from evident in marked contrast with its  
presence in a variety of theoretical models. It has been argued that 
this may be due
to the occurrence of medium-induced non-local interactions
\cite{kpzfailure}, and/or long instability-induced transients that 
mask the asymptotic KPZ scaling \cite{cuerno}. 

The Family-Vicsek scaling picture (\ref{width}) of the EW
equation also holds for the KPZ. It turns out that in one
dimension the stationary probability distribution
\begin{equation}
P(h)= \exp \Big( -{\nu \over 2D} \int (\nabla h)^2 dx \Big),
\end{equation}
which is a solution of the Fokker-Planck equation of the linear
theory ($\lambda =0$) in any dimension, is also a solution of the
Fokker-Planck equation corresponding to the KPZ. Hence, the
nonlinearity does not affect the steady state solution and consequently 
the two interfaces have the same roughness exponent $\zeta=1/2$.
The other two exponents may be obtained from the relation
$z=2-\zeta$ \cite{krug2}, valid in any dimension, and
from $\beta=  \alpha/z$, previously derived. In dimensions $d>1$,
the scaling exponents cannot be obtained analytically, nor perturbatively.
At $d>d_c=2$, both a weak and a strong-coupling regimes occur depending on
whether the value $\lambda^2 D /\nu^3$ is smaller or larger than 
a critical value. 
In the weak-coupling regime, which is unstable at the 
critical dimension $d_c=2$, $\lambda$ vanishes asymptotically  
leading to 
the EW behavior $z=2,\zeta=0$. By contrast, the strong-coupling regime 
is controlled by a fixed point that is not reached by perturbation 
analysis. 
Several conjectures for the critical
exponents have been put forward \cite{barabasi}, including a
recent claim of a mathematical proof \cite{ghaisas}. Another
controversial aspect of the KPZ behavior concerns the existence of the 
upper critical dimension. For a
more elaborated discussion of these (rather technical) matters, the 
reader is referred to existing reviews \cite{barabasi,hz,krugs}.

The KPZ nonlinearity is
related to the lateral growth that occurs when a depositing
particle sticks to the first particle it encounters on the surface
\cite{barabasi}. As a consequence, by contrast to EW-like surfaces, 
interfaces governed by the KPZ equation have a nonzero velocity at $\mu 
=0$,
\begin{equation}
v= \int_0^L \langle \partial_t h \rangle d{\bf x} = \lambda
\int_0^L \langle (\nabla h)^2\rangle d^d x = \lambda m^2,
\end{equation}
where $m^2$ is the overall quadratic slope of the interface. Since
the stationary probability distribution is known in $d=1$, $m^2= -
\partial_\gamma \ln Z$ can be computed through the simple Gaussian
path integral,
\begin{equation}
Z= \int {\cal D} h \exp \Big(-\gamma \int(\nabla h)^2 dx \Big).
\end{equation}
with $\gamma = \nu/2D$. Thus, $v=D\lambda /4\nu a$, which depends
on the microscopic scale $a$ and is therefore nonuniversal; however, 
it has a universal finite-size correction $v(L)=D\lambda /2\nu L $, a
result that holds for $t \gg L^z$. See \cite{krug} for a
different, more detailed derivation.

As the nonlinear term of the KPZ is purely kinetic
in origin, the KPZ equation cannot be obtained by differentiation of an
equilibrium Hamiltonian such as (\ref{ham}).\footnote{It has been
argued, however, that equation (\ref{kpzeq}) is not correct unless
$\partial_t h$ and $\eta$ are projected along the normal direction,
in which case a non-bound Hamiltonian exists \cite{baush}.}
Nevertheless, it can be mapped onto a linear diffusion-equation by a
simple change of variables, the so-called Cole-Hopf transform
$h({\bf x},t)= (\nu / \lambda) \ln n({\bf x},t)$, yielding
\begin{equation}
{\partial n({\bf x},t) \over \partial t} = \nu \nabla^2 n + n
\eta. \label{directedpolymer}
\end{equation}
This constitutes the {\em directed polymer representation} of the
KPZ \cite{barabasi}. We shall take advantage of this
representation in a later section but, for now, simply point out
the presence of the {\em multiplicative noise} term.

\section{BOUND INTERFACES}

Let us assume that the interfacial fluctuations are 
restricted by the presence of another interface, by a physical barrier, 
or
by any other mechanism capable of confining it. We will see
shortly how to incorporate this effect into a growth equation. Consider 
a rigid, impenetrable wall that prevents large interfacial 
excursions in a given direction. In this case, the interfacial width 
approaches a constant value,
$W(L,t) \to \xi_\bot$, as $L,t \to \infty$ even if the free interface is 
rough. Therefore, a bound interface is never rough. Moreover, since 
the fluctuations are cutoff in a given direction, there will be an 
effective
fluctuation-induced repulsion between the wall and the interface,
leading to the possibility of interfacial pinning-depinning (or
unbinding) transitions.

In the following we briefly review the behavior of a
modified EW equation describing an interface bound to a wall, for
which the aforementioned behavior can be worked out explicitly
within the mean-field approximation \cite{lipowskyfisher}. The case of
the bound KPZ interface will be tackled afterwards.

\subsection{A bound Edwards-Wilkinson equ-ation}

Restricting the height variables in the EW equation to, for
instance, positive values requires supplementing (\ref{ham})
with the {\em hard} wall condition $V(h<0)= \infty$. This is not
only inconvenient for analytical studies, but also
unnecessarily limited in scope. Therefore, more general types of
walls will be considered, the only requirements being the {\em
soft} wall condition $V(h) \to \infty$ as $h \to h_{wall}$, and
$V(h)  \to 0$ as $h \to \infty$ with $V(h)$ differentiable. 
In this case, the wall interacts with the interface, in addition to 
cutting off some of its fluctuations. The large-scale interfacial behavior 
as well as the character of the unbinding transitions are not 
expected to change as the hard-wall is replaced by a soft-wall 
\cite{lipowskyfisher}. We therefore consider the EW growth equation in the 
presence of an effective interface potential $V(h)$ \cite{lipowsky,grant}
\begin{equation}
{\partial h({\bf x},t)\over \partial t} = \nu \nabla^2h +\mu
-{\partial V \over \partial h} + \eta({\bf x},t), \label{ewwall}
\end{equation}
where $\eta$ is Gaussian white noise with the same mean and
variance as in (\ref{eweq}). The region $y > h({\bf x})$
corresponds to the stable thermodynamic phase, say $A$, the region
$y < h({\bf x})$ corresponds to a second phase $B$ that coexists with $A$ 
when $\mu=0$ and $h({\bf 
x},t)$ is the local height of the $A B$ interface measured from the wall
(see figure (1)).

$\nu$ is the interfacial tension of the $A B$ interface
and $V(h)$ accounts for the net interaction between the interfaces
binding the $B$ layer (the {\em wetting} film), in this case the
substrate and the $A B$ interface. In the simple case where
$A B$ corresponds to a vapor-liquid interface, a repulsive $V(h)$
simulates adsorption of the liquid phase from the vapor. The
form of $V(h)$ depends on the nature of the forces between the
particles in the fluid and with the wall, and its derivation
from bulk, microscopic Hamiltonians is far from trivial. Ideally
one constrains the interface, away from its equilibrium flat
position, in the configuration $h({\bf x})$ and using the
microscopic Hamiltonian, takes a partial trace over the bulk
variables. Of course this cannot be done exactly for realistic
Hamiltonians. In approximate derivations one usually requires
consistency at the mean-field level. If all the microscopic
interactions are short-ranged, one may take for sufficiently large
$h$ at bulk coexistence, \cite{bhl}
\begin{equation}
V(h)=b(T)e^{-h}+c e^{-2h}, \label{sh-potential}
\end{equation}
where $h$ is measured in units of the $B$-phase bulk
correlation length, $T$ is the temperature, $b(T)$ vanishes
linearly as $T-T_w$, the wetting temperature (see below) and $c>0$. 
If, instead, one considers
long-ranged (van der Waals) interactions, the potential has the
general form \cite{dietrich}
\begin{equation}
V(h)=b(T) h^{-m}+c h^{-n}, \qquad n>m>0, \label{lr-potential}
\end{equation}
and in this case $b(T)$ is related to the Hamaker constant. In all cases, 
if we add the contribution from the chemical potential 
difference $\mu h$ to $V(h)$, the behavior of the system  will be 
controlled by the latter term, since this is a bulk contribution that 
dominates over the surface term $V(h)$ except at bulk coexistence (figure
(2)). Finally, note that the present description
assumes the existence of an interface and thus it is only valid
below the bulk critical temperature, where distinct ``liquid'' and
``gas'' phases are defined.

Some remarks concerning the connection of equation (\ref{ewwall})
with wetting phenomena follow. Wetting occurs when the wall
preferentially adsorbs one of the phases, say $B$, while the bulk may be 
in a different thermodynamic state. At a wetting transition, 
the thickness $\langle h \rangle$ of the adsorbed $B$ layer diverges. This 
occurs at all
temperatures above a certain wetting temperature, $T_w$, at
bulk phase coexistence, i.e. at $\mu=\mu_c=0$; by contrast, at $\mu \not= 
\mu_c$ the thickness of the liquid film may be large, but cannot diverge 
(bound interface).

Equation (\ref{ewwall}) is a dynamic model for the relaxation of
the interfacial height $h({\bf x},t)$ towards the equilibrium
configuration that minimizes the Hamiltonian
\begin{equation}
{\cal H}(h)=\int_0^\infty d{\bf x}\Bigg[ {\nu \over 2}  (\nabla
h)^2 +V(h)\Bigg] \label{hamwall}
\end{equation}
in the limit $t \to \infty$. Within mean-field approximation
$\langle h \rangle$ follows from the condition $\partial V(h) /
\partial h=0$, whereby one finds that the equilibrium thickness of
the wetting layer for an attractive wall ($b<0$) and long-ranged
forces with $m=3, n=2$, (non-retarded van der Waals forces
\cite{dzyaloshinskii}) is given by $\langle h \rangle =-2b/3c$. A
{\em critical wetting} transition takes place as $b \sim T-T_w \to
b_w=0^-$, i.e. the wetting layer thickness diverges as $\langle h
\rangle \sim |b-b_w|^{\beta_c}$ with $\beta_c=-1$. This mean-field result 
was observed experimentally
\cite{meunier_lrcw} since the upper critical dimension of the long-ranged  
system is less than three \cite{lipowsky:prl84}.

Critical wetting driven by short-range forces may also occur
since the long-ranged interactions can be neglected if 
the bulk correlation length is sufficiently large i.e., when $T_w$ is
close to the bulk $T_c$. This has been confirmed recently by the
observation of effective short-range critical wetting at the
liquid-vapor interface of methanol-alkane mixtures \cite{nature}.
Within mean-field theory, $\langle h \rangle =\ln (-2c/b)$ and,
consequently, $\beta_c= 0 (log)$. However, the measured exponents,
which are also mean-field like, are at odds with the theoretical,
renormalization-group-based predictions \cite{nature}.

By contrast, the wetting transition may be driven by the chemical
potential difference between the $A$ and $B$ phases, at
any temperature above $T_w$; this is always a continuous
transition and it is known as {\em complete wetting}
\cite{dietrich}. A study of complete wetting transitions requires
adding a linear term $\mu h$ to the Hamiltonian, where $\mu$ is
the chemical potential difference of phases $A$ and $B$. Thus, on
approaching coexistence for $T>T_w$ ($b>0$ at the mean-field
level), $\langle h \rangle$ diverges as $\langle h \rangle \sim
|\mu-\mu_c|^{-\beta_h}$, with $\mu_c=0$, and $\beta_h = 0(log)$ and
$\beta_h=1/(m+1)$ for short- and long-ranged forces,
respectively. Experimental observations of complete wetting
transitions are numerous and are characterized, in general, by
long-range mean-field exponents (at least far from the bulk
critical temperature). The interested reader is referred to
\cite{dietrich} for reviews, and to \cite{reviewbonn} and
\cite{reviewmc} for a detailed account of recent experimental and
Monte Carlo results, respectively.

This rich behavior is summarized in the schematic phase diagram of
figure (3): (along path 1) $\langle h \rangle$ diverges 
continuously as coexistence is approached from the gas phase 
(complete wetting); (path 2) as $T$ approaches $T_w$ at
coexistence $\langle h \rangle$ may diverge either discontinuously
(first-order wetting ) or continuously (critical wetting);
(path 3) the wall remains non wet when
coexistence is reached . Clearly, in
the wetting regime at long times, the form of $V(h)$, determines
the rate of growth that is logarithmic, i.e. $\langle h \rangle
\sim t^{\theta_h} \sim \ln t$ or $\theta_h=0$ for short-ranged 
interactions and a
power law, $\langle h \rangle \sim t^{1/(2+m)}$, for
long-ranged repulsive interactions decaying as $h^{-m}$
\cite{lipowsky}. Off coexistence, for a small and negative 
$\delta \mu \equiv \mu -\mu_c$,
the wetting layer thickness grows with time as predicted by the mean-field
theory, provided that $t$ is much smaller than the correlation time 
$\delta \mu^{\nu_t}$. Then, it relaxes exponentially to its equilibrium value
$\langle h(t=\infty) \rangle$. These exponents are related by 
$\theta_h =\beta_h/z \nu_x$, and $\nu_t =-2\nu_x$ \cite{lipowsky}. 
On the other hand,  if $\delta \mu$ is positive, $\langle h(t) \rangle$ grows 
linearly. The results for the exponents are summarized in
table (\ref{table1}) and Monte Carlo simulations of the growth of wetting
layers can be found in \cite{mon,ebner}.

The behavior below $b_w$ deserves some comments. 
Wetting only occurs at coexistence.
The first-order transition that takes place at
$\mu =0$ has to be
distinguished from wetting because it is not driven by the
substrate. In fact, the wall is not wet at the transition
point and the depinning transition that occurs upon crossing the boundary
line $\mu=0$ is, as such, trivial since to the right of
the $\mu=0$ line liquid is the only stable thermodynamic phase,
with or without a wall. 

Lastly, we comment briefly on how the effective wall interaction
due to fluctuations comes about. As a result of the confinement of
the interface, there is an increase in the elastic bending energy
and an entropy loss. For $d=1$ both contributions are of the same
order and can be estimated as $\Delta e \approx T \Delta s =
\Delta V \sim  h^{-2}$, while for $d=2$ one has $\exp(-2h^2)$
\cite{lipowskyfisher}. By simply adding $\Delta V$ to $V(h)$
one gets a description of the effects of fluctuations within
mean-field theory.

\subsection{A bound Kardar-Parisi-Zhang e-quation}

Consider the following dynamic interfacial model
\begin{equation}
{\partial  h({\bf x},t) \over \partial t} = \nu \nabla^2 h + \lambda
(\nabla h)^2 +\mu -{\partial V(h) \over \partial h} + \eta({\bf x},t).
\label{bkpz}
\end{equation}
The fact that the KPZ equation is not invariant under the
transformation $h \to -h$ implies that the sign of $\lambda$
assigns an {\it orientation} to the wall, distinguishing two types:
{\em upper} and {\em lower} walls. An upper (lower) wall
restricts (or even prevents) large interfacial excursions into the 
region $h > 0$ ($h<0$). Thus, for negative values
of $\lambda$ the interface is pushed on average against a lower
wall, while for positive $\lambda$ it is pulled away from it, and
exactly the opposite occurs at an upper wall. As will be shown
below, for a given sign of the non-linearity upper and lower walls
lead to quite different phenomenologies, but the case $\lambda >0$
and a lower wall is completely equivalent to $\lambda <0$ and an
upper wall \cite{munozhwa}. We stress that, were it not for the
presence of the nonlinear term, such distinctions would not have
been necessary (both equilibrium wetting and dewetting are
symmetric phenomena).

As noted earlier, in the absence of the limiting wall, bulk
coexistence no longer obtains at $\mu=0$. Rather, a nonzero
chemical potential $\mu_c \sim \lambda$ is required to balance the
force exerted by the nonlinear term on the tilted regions of the
interface. For $\lambda =0$ the model reduces to the equilibrium
one and $\mu_c=0$ as usual. As in the previous section, we
distinguish the cases of short-- and long--ranged interactions. Also,
for definiteness, in what follows the wall will always be assumed
to be a lower one, while $\lambda$ may take either sign.

\subsubsection{Short-ranged forces}

Consider the Langevin equation (\ref{bkpz}) where $V(h)=b
e^{-ph}+c e^{-2ph}$. The parameter $p$ controls the nature of
the wall: the limit $p \to \infty$ corresponds to an impenetrable
wall and the sign of $p$ determines whether it is an upper (-) or
a lower one (+). The same equation with $c=0$, $b>0$ (repulsive
wall), and $\lambda<0$ was first studied by Tu {\em et al.} 
\cite{MN2}, and by Mu\~noz and Hwa
\cite{munozhwa} for arbitrary $\lambda$ in the context of nonlinear 
diffusion with multiplicative noise (see later in this article). The effect of
inert and attractive walls was then investigated by Hinrichsen
{\em et al.} in \cite{haye1,haye2} using a discrete solid-on-solid
model with dynamics that violate detailed balance. The model includes 
particle adsorption and desorption rates, plus an
additional growth rate at the wall that simulates a short-ranged
attractive or repulsive interaction between the wall and the interface. 
Their results were subsequently
confirmed and expanded by Giada and Marsili \cite{marsili} who
mapped (\ref{bkpz}) to a Langevin equation with multiplicative
noise and analyzed it within mean-field, and by ourselves
\cite{nos,lisboa}, by means of direct integration of
(\ref{bkpz}) with $\lambda <0$. The work of \cite{marsili}
privileges the role of the noise as the driving force, while in
\cite{nos,lisboa} the noise strength is fixed and $b$ and $\mu$
are taken as control parameters. The case $\lambda >0$ was
recently investigated in \cite{brazilian}.

Some of the exponents defined 
in table (\ref{table1}) may be
related to KPZ exponents. In particular, the dynamic exponent
$z$ is unchanged from its KPZ value, and $\nu_x=1/(2z-2)$ for
any value of $\lambda$ \cite{MN1,MN2}. Thus, in one dimension 
$\nu_x=1$ since $z=3/2$. The two remaining exponents, $\beta_h$
and $\theta_h$ are related by the scaling form 
$\theta_h=\beta_h/ \nu_x z$.
 
The long-ranged repulsion exerted by the wall on the interface, mentioned 
previously, may be estimated using a simple scaling
argument: the wall makes itself felt when the characteristic distance of 
the interfacial excursions is of order $h$, $\xi_\bot \sim h$. From 
(\ref{width}) 
and the definition of $\nu_x$, we find  
$\xi_\bot \sim \xi ^{\zeta} \sim \delta \mu ^{-\nu_x \zeta}$, leading
to an effective repulsive force $h^{-1/\nu_x \zeta} \sim h^{-2/\nu_x}$, 
since $\zeta=1/2$ in $d=1$ for both KPZ and EW. By substituting
$\nu_x =2/3$  (see table (\ref{table1})) we recover the EW result, while 
$\nu_x=1$ yields $V(h) \sim h^{-1}$ for the KPZ \cite{munozhwa}.

\underline{$\lambda <0$.} We first consider a negative
KPZ non-linea\-rity. Before discussing the results, we note that upon
inverting the sign of $h$, $\lambda$  changes to $- \lambda$ and
$p$ to $-p$, so that the exponential now acts as an upper wall and
$\lambda >0$ pushes the interface against it. This shows that the
cases $\lambda >0$---upper-wall and $\lambda<0$---lower-wall are
equivalent.

The associated phase diagram in the $b-\mu$ plane is depicted in
figure (5). The solid line is the continuous phase
boundary between depinned and pinned phases. Between the dashed
lines, both of which correspond to first-order boundaries, the
pinned and depinned phases coexist as stationary solutions of the
dynamical equation. The three lines meet at the tricritical point
$(\mu_c,b_w)$.

The unbinding transition at $\mu=\mu_c$ for any $b>b_w$ is the
analogue of the equilibrium complete wetting phase transition
(path 1 of figure (4)). The exponent governing the
divergence of $\langle h \rangle$ is $\beta_h \approx 0.41$, with
error bars that exclude the equilibrium value $\beta_h = 1/3$
\cite{haye1,grasem}. Along path 2, which corresponds to a
nonequilibrium critical wetting transition, $b$ is progressively
increased until the nonwet phase becomes unstable at $b_w$. It is
found that the critical temperature is depressed from its
mean-field value $b_w=0$, with an associated critical exponent
$\beta_c \approx -1.2$ \cite{lisboa}. 

As for the behavior below $b_w$
(attractive wall) there are two first-order depinning transitions
by varying $\mu$  (paths 3 and 4 of figure (4)). Recall that,
if the wetting temperature is to be a meaningful concept, the
depinning transition for $b<b_w$ cannot be viewed as a wetting
transition or, in other words, these transitions occur not on
approaching coexistence, but when {\em crossing} the boundary
line. For sufficiently negative values of $b$ the pinned phase
becomes unstable at $\mu^*(b)$ (path 4), and the depinned phase at $\mu_c <
\mu^*(b)$ (path 3). Consequently, in the range $\mu_c < \mu < \mu^*(b)$
both phases coexist in the sense that if the interface is
initially close to the wall it remains pinned, while if it is
initially far from the wall it detaches and moves away at a
constant velocity. As a consequence of the ``broad'' phase
boundary there is the possibility of defining critical wetting
along a range of different paths, delimited by the dashed lines of
figure (4). It has been checked that the exponent
$\beta_c$ does not change when the tricritical point is approached
along different paths within this region \cite{lisboa}.

The fact that the coexistence region is finite rather than a line
is a nonequilibrium effect since at equilibrium two-phase
coexistence only occurs on a surface of dimension one less
than the dimension of the space of thermodynamic parameters. An 
equilibrium system initially in a state other than the equilibrium one
will be destabilized by local fluctuations in the form of 
droplets of the equilibrium phase. To ensure
generic multistability, a robust mechanism for eliminating these
droplets of the minority phase must exist. In the present
case, such a mechanism was found by Hinrichsen {\em et al.}
\cite{haye2}: when, due to thermal fluctuations, a segment of the
interface overcomes the potential barrier and gets out of the
potential well (see figure (2)), it rapidly
acquires a triangular shape, after which it shrinks at a constant
velocity driven by the negative nonlinear term, in a time proportional
to its size. Typical interface profiles resulting from numerical
solutions of (\ref{bkpz}) in $d=1$ and 2 are shown in figure
(5) (similar configurations are described in the
discrete model of reference \cite{haye2}). The largest size of the
depinned regions, i.e. the size of the triangular bases,
increases as the instability threshold $\mu^*$ is approached. Once
the last site has detached, the interface ``takes off'' and a
transition from a pinned to a moving phase takes place. In fact,
owing to finite-size effects the only stable phase within the
coexistence region is the moving one, making the analysis of
the pinned regime rather hard. This difficulty is circumvented by
studying $\tau$, the time taken by the interface to depin in the limit $L
\to \infty$. For $\mu > \mu^*$, $\tau$ saturates with
increasing system-size and thus the interface detaches in a finite
time. For $\mu_c < \mu < \mu^*$, however, $\tau$ grows
exponentially with the system-size so that the pinned phase
becomes stable in the thermodynamic limit. The stationary
distribution of triangles as a function of their base-size turns
out to be exponential. This indicates that there is a typical size
for the depinned regions and rules out the possibility that growth
is driven by a coarsening mechanism \cite{lisboa}. Another
interesting feature is that, for a given pair of parameters $\mu$
and $b$, all the triangular facets have the same slope, $s$, which
can be determined through the relation $|\lambda_R| s^2 = \mu$,
where $\lambda_R$ is the renormalized nonlinear coefficient of the
KPZ. Historically, the possibility of a finite phase boundary was
first discussed in Toom's north-east-center voter model
\cite{toom}. Other examples include systems of
harmonically-coupled, identical nonlinear constituents under the
simultaneous influence of additive and multiplicative noise
\cite{muller}, and a Leshhorn automaton subject to transient
velocity-dependent forces \cite{jennifer}. The relation between
these systems and condiditons for generic phase coexistence 
are discussed in \cite{coexistence}.

It has been pointed out that upon diminishing both the depth of
the potential well and the surface tension, the first-order phase
transition at $\mu^*$ becomes continuous with {\em directed
percolation} (DP) critical exponents \cite{synchro}.\footnote{To
observe DP behavior a different order-parameter, namely the
density of pinned sites, has to be used. More on this later} 
This is in line
with a recent claim that first-order phase transitions in
one-dimensional nonequilibrium systems with fluctuating ordered
phases are impossible, provided there are no conservation laws,
long-ranged interactions, macroscopic currents, or special boundary
conditions \cite{haye_firstorder}. If this is the case, the reported 
first-order behavior will be a very long transient effect. 
The presence of DP behavior is now a well-established
result that was found in a variety of discrete models
\cite{synchro,lipowski_droz,ginelli}.
Nevertheless, we cannot discard the presence of a tricritical point for
sufficiently negative values of $b$, separating DP from first-order
transitions. At present, this question remains open.

\underline{$\lambda > 0$.} Consider $\lambda >0$. Early
results were reported in \cite{munozhwa}, but the observed critical
exponents are far from their true asymptotic values. Later studies
revealed marked differences when compared to the system with a
negative nonlinearity and attractive walls
\cite{haye2,brazilian}. These are apparent in the phase diagram of
figure (4), similar to that of $\lambda =0$ (figure
(3)). Before we proceed, we note that
the results of this section were obtained using Monte Carlo simulations of
discrete models. Attempts at numerical integration of a bound
KPZ equation failed to reach a stationary state. It is well known 
that the results from numerical integration may disagree with the
predictions from the continuum KPZ \cite{KPZproblem}. The usual
way to handle the integration of the KPZ equation
with short-ranged forces and negative nonlinearity, the Cole-Hopf
transformation, is also plagued with instabilities. 
A mean-field analysis, however, exits for
the oder-parameter $n=\exp(-h)$ (see section 4).

We start by describing the results for a repulsive wall. A continuous
wetting transition is found as $\mu \to \mu_c$ when $b>b_w$
($b_w=0$ at the mean-field level; path 1 of figure (4)).
The behavior of the order parameter in the vicinity of the critical point
yields $\beta_h = 0.52(2)$, and the time evolution of $\langle h
\rangle$ at $\mu_c$ behaves as $\langle h(t) \rangle \sim
t^{\theta_h}$, with $\theta_h = 0.355(15)$ \cite{brazilian}. 
From the scaling relation $z=\beta_h/\nu_x \theta_h$,
$z=0.52(2)/0.355(15)=1.5(1)$ in good agreement with the theoretical 
prediction $z=3/2$.  
These results, however, disagree
with those of reference \cite{haye1} probably due to the presence of 
extremely
long transients in the latter. The results are collected in table 
(2).

For attractive walls, we note that there is a line of
first-order depinning transitions ending at a tricritical point
at $b_w$ (path 3 of figure (4)). By contrast to what happens
for $\lambda <0$, phase-coexistence is restricted to a line since 
the mechanism responsible for the elimination of droplets of the depinned 
phase works 
only for negative $\lambda$. Finally, the tricritical behavior 
associated with the transition along path 2 has not been investigated yet.

\subsubsection{Long-ranged forces}

Next, consider the KPZ equation in the presence of a long-ranged 
potential $V(h)=b(T) h^{-m}+c h^{-n}$. The resulting equation is not 
amenable to analytical treatment and therefore we are limited to numerical 
results from Monte Carlo simulations of one-dimensional, discrete models. 
Recently, Lipowski and Droz \cite{lipowski_droz} 
studied an interfacial growth model
with dynamics inspired by the synchronization transition in coupled map
lattices and included a power-law interaction between the interface
and the wall.\footnote{The relationship between
synchronization transitions and a bound KPZ will be seen in the next 
section.}
The model, however, lacks a surface tension and its microscopic 
rules are such that it is difficult to ascertain whether it may 
be described by a bound KPZ equation. 

Interestingly enough, the KPZ nonlinearity appears to be irrelevant
above the wetting temperature $b_w$ for any sign of $\lambda$,
and an equilibrium
complete-wetting transition is found along paths 1 of 
figure (4) \cite{lipowsky_kpz}. 
The associated exponents are collected in table (\ref{table1}) and have
been verified numerically in a simulation of a variant of a model
introduced in \cite{krugs}, whose continuum counterpart is known
to be the KPZ with $\lambda <0$ and a lower wall \cite{sem_publicar}.

Below $b_w$, a phase diagram similar to that for short-ranged interactions
is found (figure (4)) for either sign of $\lambda$. In particular, for
$\lambda <0$ and for the set of parameters explored, the transition
at $\mu^*(b)$ (path 4) is characterized by DP exponents when the 
order parameter $n=\exp(-h)$ is used \cite{sem_publicar}.
For $\lambda >0 $, to our knowledge, there are no published results. 
Results of ours indicate a phase diagram similar to 
the one shown on the right of figure (4): along path 3 the
system undergoes a discontinuous transition. The analogue to
a nonequilibrium critical/first-oder wetting transition along path 2
has not been investigated.

\section{RELATED PHENOMENA}

In this section, the directed-polymer representation of the KPZ is
exploited to analyze bound KPZ interfaces. As shown in section
2.2, the KPZ can be linearized by means of the Cole-Hopf transformation, 
$n({\bf x},t) = \exp[\alpha h({\bf x},t)]$, at the cost
of introducing a multiplicative-noise (MN) term. This transforms
the KPZ equation (\ref{kpzeq}) into 
$$ {\partial n({\bf x},t)
\over \partial t} = \nu \nabla^2 n + n \eta. \eqno(12) $$
Note that the noise amplitude vanishes for $n=0$, and neither
the deterministic dynamics, nor fluctuations can take the system
out of this state. For this reason the configuration $n=0$ is
known in the MN literature as an {\em absorbing} state, while
$n\not= 0$ is commonly referred to as the {\em active} phase. This
representation has been successfully applied to the case of
short-range interactions with a negative non-linearity, enabling
analytic calculations both at the mean-field level
\cite{marsili,nos,lisboa,muller}, and beyond via a renormalization
group approach \cite{MN2,MN1}. Let us consider again (\ref{bkpz}) in
the presence of a short-ranged potential with, for simplicity, 
$-\lambda = \nu=D=1$ (different coefficients for the
Laplacian and the KPZ nonlinearity may be accounted for by a proper
choice of $\alpha$). The change of variables $h=-\ln n$ leads to
\begin{equation}
{\partial n({\bf x},t) \over \partial t} = \nu \nabla^2 n -
{\partial V(n) \over \partial n} + n \eta, 
\label{mn}
\end{equation}
with $V(n)=\mu n^2/2+bn^3/3+cn^4/2$, and where we have made use of
the Stratonovich calculus \cite{vankampen,gardiner}.\footnote{The
difference between the Ito or the Stratonovich
results is a trivial shift in $\mu$.} In general, potentials
of the form $V(n)=b n^{p+2}+cn^{2p+2}$ with $p>0$ result in
equivalent effective Hamiltonians since, when expressed in terms
of $h$, $p$ can be eliminated by defining the length scale. In the $n$ 
language, unbinding from the wall, $\langle h \rangle
\to \infty$, corresponds to a transition to an absorbing state,
$\langle n \rangle \to 0$, recovering the usual behavior
of Landau theory where the order parameter
becomes small rather than large near the
transition.\footnote{Strictly speaking, at unbinding the system is in a 
state that evolves continuously to the absorbing state.} To complete the 
analogy between the MN and the interfacial descriptions, 
one can give a physical interpretation to $n$ by noting that it is
the density of sites at zero height, $n({\bf x},t)
=\delta_{h({\bf x},t),0}$ \cite{haye3}. 
The behavior of $\langle n \rangle$ is characterized by 
a set of exponents similar to that of $\langle h \rangle$:
$\langle n \rangle \sim t^{-\theta_n}, 
\langle n \rangle \sim \delta \mu^{\beta_n},
\xi \sim \delta \mu^{-\nu_x}$, and $\xi \sim t^{1/z}$.
Some of these exponents are obtained from the KPZ exponents 
using scaling arguments. For instance, 
it was shown in \cite{MN2,MN1} that the dynamic exponent for
the MN case is identical to the value of the KPZ $z$, that
$\nu_x$ is related to it through $\nu_x=1/(2z-2)$,
and that $\beta_n>1$. Also, the scaling relation 
$\theta_n = \beta_n/\nu_x z$ is satisfied. 
It is therefore not surprising that the phase diagram
of the MN equation (\ref{mn}) is very similar to that of the
KPZ. There is a strong noise phase for $d\leq d_c=2$, and both
a weak and a strong noise phases for $d>2$ depending on the 
noise intensity $D$. Weak and strong phases obtain for values of 
$D$ that are, respectively, smaller and larger than a critical value
\cite{MN2}. See \cite{genovese} for a comparative list of critical
indices below and above $d_c$.  

In this context,
(\ref{mn}) is a Langevin-like equation for a reaction-diffusion
process where $n$ is a coarse-grained particle density. Indeed, it
has been argued in \cite{tauber} that (\ref{mn}) is the
equation governing the pair contact process with diffusion, $2A
\to 3A, 2A \to \emptyset$, the critical behavior of which is
currently under debate (see \cite{haye3} and references therein).
A different realization of (\ref{mn}) comes from the field of
synchronization in spatially extended systems. Pikovsky {\em et
al.} \cite{pikovsky} established that the difference field,
$n({\bf x},t)$, between two coupled-map lattices follows
(\ref{mn}) with $c=0$. Two replicas of a coupled-map lattice are
synchronized when $n=0$, $n\not= 0$ corresponding to the asynchronous
regime, and the active-absorbing, or the pinning-depinning, phase
transition is analogous to a synchronization transition. As a
last example, (\ref{mn}) has been studied in \cite{raul} in the
context of spatio-temporal intermittency. The unsuspected
connections between these problems are illustrated in what
follows.

The mean-field approximation to (\ref{mn}) consists of
discretizing the Laplacian as $1/2d \sum_j (n_j-n_i)$, where
$n_i=n(x_i,t)$ and the sum is over the nearest-neighbors of
$i$. Substituting the values of the nearest-neighbors by the
average field, $\langle n \rangle$, a closed Fokker-Planck
equation is obtained for $P(n,t,\langle n \rangle)$. 
This approach takes into account the effect of the 
noise and of a spatially varying order parameter. The
steady-state solution is then found from the self-consistency
requirement
\begin{equation}
\langle n \rangle = {\int_0^\infty dn \ n P(n,\langle n \rangle)
\over \int_0^\infty dn \ P(n,\langle n \rangle)},
\end{equation}
resulting in a phase diagram equivalent to that of figure
(4) with a tricritical point at $(\mu=0, b_w=0)$
\cite{nos,lisboa}. Thus, for repulsive walls a complete wetting
transition characterized by~$z=2, \nu_x=1/2$, and $\theta_n=1$, is found 
at $\mu_c=0$ for any $b>0$.
Different mean-field approaches yield different results for the
exponent $\beta_n$, namely $1/p$ \cite{genovese} and  $D/2$ 
\cite{marsili},
but this has been clarified in \cite{birner} where a crossover from  
$1/p$ to a nonuniversal, continuous exponent $D/2\nu$ was identified. 
For attractive walls, the three regimes previously described are 
recovered,
namely, an active phase for $\mu <\mu_c=0$, an absorbing one for
$\mu > \mu^*(b)$, and phase coexistence of the active and 
absorbing phases in the region $0 < \mu < \mu^*(b)$. A
qualitatively equivalent picture is obtained using a different
space of parameters, for instance the strength of the noise vs.
$\mu$ \cite{marsili}.

Beyond mean-field theory, (\ref{mn}) is known to be
superrenormalizable above $b=0$, i.e. the Feynman graphs may be
computed to all orders and resummed \cite{MN2,MN1}. This does not
imply that all the critical exponents are given, as the
renormalization group-flow equation has runaway trajectories
that are supposed to converge to the strong coupling fixed point.
Representative results for $d=1$ are summarized in table (2).

The rich phenomenology obtained for attractive walls is also observed 
within the MN framework. Active and absorbing phases coexist over
a finite area $\mu_c < \mu < \mu^*(b)$, and lose their stability at,
$\mu^*(b)$ and $\mu_c$, respectively. The 
phase-coexistence regime may be characterized by analyzing the single-site 
stationary 
density function, defined as the average of $n(t)$ over pinned states 
rather than over all runs. This is depicted in figure (6) 
and shows how the histogram develops a maximum at $n=0$ as $\mu$ 
approaches the stability edge $\mu^*(b)$, beyond which it changes abruptly 
into a delta function and the pinned phase becomes unstable. The 
simultaneous presence of two-peaks indicates that a fraction of the 
interface depins. This is clearly seen in the space-time snapshot
of a numerical solution of (\ref{mn}), on the left of figure 
(6). The main features of such a pattern,
spontaneous formation of domains with a wide range of sizes and lifetimes,
has been identified as distinctive of spatio-temporal intermittency 
\cite{raul}.

Numerical simulations provide evidence of a first-order phase transition
at $\mu_c$. As was pointed out previously, both 
first-order and continuous transitions are observed at $\mu^*(b)$ 
depending on the model parameters. For instance, a continuous DP-like 
transition was recently reported in \cite{synchro} as the numerical 
solution of (\ref{mn}) with $b=-9, c=8, \nu=0.1$, and $D=1$. By contrast, 
for $b=-3, c=2, \nu=D=1$, a first-order transition was found. The 
minimal Langevin equation that describes the DP universality class is
\begin{equation}
{\partial n({\bf x},t) \over \partial t} = \nu \nabla^2 n 
-\mu n - b n^2 + \sqrt{n} \eta, 
\label{dp}
\end{equation}
which differs from (\ref{mn}) by the $n^{1/2}$ noise amplitude. Thus, it 
appears that DP-like noise takes over the MN-noise, proportional to the 
field $n$, as the wall changes from repulsive to attractive in a given
range of parameters. Deriving (\ref{dp}) from (\ref{mn}) is an interesting 
theoretical problem that remains open.

The similarity of behavior of KPZ interfaces bound by short-- and 
long--ranged substrates also shows up in the framework of the order
parameter $n$. Recall that an interface bound to the substrate is 
described by a particle-density field with low (high) density segments
corresponding to detached (pinned) interface domains. 
Interface fluctuations can then be associated with the creation, 
annihilation, or merging of clusters of particles (see figure (7)). 
In particular, the disappearance of triangles corresponds to a 
cluster-coarsening whose
dynamics, being ultimately controlled by DP exponents, can be modeled
by a contact-process. Recently, a lattice model of a generalized contact 
process with long-ranged interactions between the edges of low-density 
segments has been investigated \cite{longranged_dp}. A 
transition in the DP universality class is found for forces that decay
sufficiently slow, and a first-order transition otherwise. Clearly, in
terms of $h$ this translates into a long--ranged interaction
between the vertices forming the triangle bases, and it is reasonable
to assume that, in turn, an effective long--ranged attraction between
the substrate and the interface must obtain. Consequently, both short-
and long-ranged interactions yield the same behavior below $b_w$
\cite{sem_publicar}.

A direct application of the Cole-Hopf transformation $n= \exp(-h)$ to
(\ref{bkpz}) with $\lambda >0$ in the presence of short-ranged
interactions yields the Langevin equation
\begin{equation}
{\partial n({\bf x},t) \over \partial t} = \nu \nabla^2 n 
-2{(\nabla n)^2 \over n} -\mu n -b n^2 -2c n^3 +n \eta, 
\label{mn2}
\end{equation}
which is (\ref{mn}) plus the extra term $-2(\nabla n)^2/n$. Apart 
from the factor -2, equation (\ref{mn2}) is the Cole-Hopf transform of 
\begin{equation}
{\partial h \over \partial t} = \nabla^2 h +\mu + b e^{-h} + c e^{-2h} +\eta
\end{equation}
which is the equilibrium EW model in the presence of a binding wall.
Note that the factor of 2 in (\ref{mn2}) cannot be absorbed
by reparametrization. Indeed, equation (\ref{mn2}) does not admit 
proper mean-field solutions, it is not amenable to standard perturbative 
field theoretical methods, and cannot be integrated numerically  
\cite{brazilian}.
Nevertheless, (\ref{mn2}) may be studied within the framework of
active/absorbing phase transitions by simulating discrete 
growth models, and monitoring the order parameter $n=\exp(-h)$.
This was done in \cite{brazilian} for two different models argued 
to correspond to a bound KPZ with $\lambda > 0$
and short-ranged interactions. Results suggest the existence of a
new universality class characterized by 
$\beta_n =0.33, \theta_n \approx 0.22, z=1.5, \nu_x \approx 1$.
Ginelli et al. carried out a mean-field analysis
based on a lattice model for KPZ-like growth, the single-step model,
that does not completely disregard fluctuations \cite{skewed}. 
They found a surface exponent $\theta_n = 1/3$ and succeeded in 
reproducing qualitatively the main features of the system 
for both signs of $\lambda$.
Finally, for a sufficiently attractive wall, the transition becomes first-order. 

$h \ or \ n?$ The Cole-Hopf transform puts the problem of bound KPZ 
interfaces in a rather different context. For
$\lambda <0$, a more compact equation, instability-free and amenable 
to a mean-field analysis,
is obtained, whereas for $\lambda >0$ the presence of $(\nabla n)^2/n$ 
suggests that the interface language is most natural. We stress, 
however, that the form of the continuum equation
does not determine the ``correct'' oder-parameter: In fact, 
proper scaling behavior is found for both $h$ and $n$ in discrete models.
DP is a counterexample 
since the interface representation of the contact
process shows clear signs of anomalous scaling \cite{interface_dp}. 
Another interesting point concerns the $n$-representation of the EW 
problem,
that is, the equilibrium limit ($\lambda=0$) of (\ref{bkpz}).
The Cole-Hopf transform fixes the ratio $\lambda/\nu$ and thus
the $\lambda=0$ limit cannot be taken when the transformation is used.
A comparison of published results on equilibrium and 
nonequilibrium
wetting, reveals that the EW plus a wetting potential 
is not equivalent to the KPZ in the weak-coupling regime plus the 
same wetting potential.

\section{EXPERIMENTAL REALI\-ZA\-TIONS}

Since the idea of a wetting transition was introduced in 1977 by 
Cahn \cite{cahn}, much experimental effort has gone into its 
realization, ultimately awarded by the confirmation of a large
number of theoretical predictions. The obvious question then
arises of what are the experimental realizations of the phenomenology 
described for the bound KPZ interface. 
As was discussed in section 2.2, the KPZ non-linearity models 
lateral growth. Although this mechanism
is unlikely to be relevant in simple fluids, it may be important
in describing systems with anisotropic interactions for which the
growth of tilted interfaces depends on their orientation. For
instance, it has been shown that crystal growth from atomic beams
when desorption is allowed is described by the KPZ equation
\cite{villain}. Furthermore, a renormalization group analysis reveals
that the KPZ term is always generated, except when excluded by
symmetry, whenever elastic objects depin in the presence of
anisotropy \cite{wiese}. This suggests that an answer may come 
from the field of crystal growth. 
Crystals grown by deposition of material onto a crystal substrate 
from a vapor phase can display wetting behavior.
At sufficiently low temperatures, the adsorbate that intervenes 
between the substrate and the vapor may form small crystallites,
which melt upon increasing the temperature, leading to the formation of 
liquid droplets. On approaching the wetting temperature, the liquid
spreads over the substrate and coats it uniformly. Growth of solid
phases is also possible, but in this case the wetting layer is likely to 
be under stress due to the misfit between the lattice constants
of the substrate and the adsorbate. Two well-known relaxing mechanisms
are the formation of misfit dislocations and the deformation of
the surface (Asaro-Tiller-Grinfeld instability \cite{review_crystal_growth}).
Faceting and, in particular, the appearance of pyramidal structures
similar to those of figure (5)
have been reported in simulations of wetting-layers and island formation in
heteroepitaxial growth \cite{biehl}. These structures can either form
directly on the substrate (Vollmer-Weber growth) or on top of a wetting-layer
of finite thickness (Stranski-Krastanov-like growth). Clearly,
further work is needed before a connection between crystal growth and 
a bound KPZ can be substantiated.

A second direction of research is the experimental verification 
of synchronization transitions in extended, one-dimensional systems.
The order parameter for such transitions is the difference 
between two dynamical systems coupled to each other, 
$n(x,t)=|u_1(x,t)-u_2(x,t)|$. Starting from different initial 
conditions, the systems become synchronized in the stationary limit, 
$\langle n \rangle=0$, above a certain critical coupling
strength, and asynchronized, $\langle n \rangle > 0$, otherwise.
The character of the transition depends on whether the largest Lyapunov 
exponent of the system, $\Lambda$, is zero or negative. In the former case, 
the transition is in the MN universality class, while in the latter
it is DP-like or discontinuous depending on the system details. Equation 
(\ref{mn}) is a stochastic model for the difference field $n$, 
with $\Lambda =\mu_c$ at the transition. Examples of such chaotic 
coupled-systems include semiconductor laser arrays and 
liquid crystals describable by the anisotropic
Ginzburg-Landau equation, for which the aforementioned behavior 
should be experimentally detectable \cite{pikovsky_book,ahlers}.
This goes hand in hand with finding a system which reproduces
the critical exponents of directed percolation, a problem that
so far has defied all attempts of solution despite extensive
empirical efforts \cite{haye5}.

The other large class of bound KPZ ---short-ranged forces, positive 
non-linearity, and a lower wall--- should also describe actual nonequilibrium 
wetting phenomena. There is nonetheless one other possible experimental
realization. Sequence alignment is a powerful tool for determining 
relatedness between sequences of proteins or DNA segments.
Algorithms have been devised to detect such correlations, the Smith-Waterman 
being among the most sensitive for finding related sequences in a 
database \cite{smith_waterman}. 
Scores are assigned to each character-to-character comparison:
positive for exact matches and negative if the two are different, or
if pairing of an element with a gap occurs. An optimal
alignment is one with the maximum possible score. Recently, Hwa {\em et al.}
have mapped this algorithm to a KPZ with a lower wall, where the location
of the critical point corresponds to the choice of the optimal scoring 
parameters \cite{hwa}.

Acknowledgements.
We are grateful to M.A. Mu\~noz for a critical reading of the 
manuscript.

\newpage 

{\center\LARGE{TABLE CAPTIONS}}
\vspace{1cm}

\noindent
Table 1: Upper critical dimensions and critical exponents in
the mean-field (MF) and the fluctuation regimes for complete
wetting transitions with short and long-ranged forces. At the
mean-field level, the exponents for short-ranged interactions 
are recovered in the $m \to \infty$ limit of the long-ranged
ones.
\vspace{1cm}

\noindent
Table 2: Summary of the universality classes discussed in the
text for short-ranged forces. For transitions other 
than DP, the theoretical predictions are $z=3/2, \nu_x=1$, 
and $\beta_n>1$ if $\lambda < 0$.

\newpage

\begin{center}
\LARGE{TABLE 1}
\end{center}
\vspace{5cm}

\centerline{
\begin{tabular}{|c||c|c||c|c||} \hline
\rule[6mm]{0mm}{1mm} {\it Exponent } & \multicolumn{2}{|c||}{\it
Short-range forces} & \multicolumn{2}{|c||}{\it Long-ranged forces}
\\ \cline{2-5}
&\rule[4mm]{0mm}{1mm} MF & $d<d_c=2$ & MF &$d<d_c(m)={2m\over2+m}$
\\
\hline \hline $\theta_h$, $\langle h \rangle \sim t^{\theta_h}$ &
0 & 1/4  & 1/(2+m)      & (2-d)/4    \\ \hline $\nu_x$, $\xi \sim
\delta \mu^{-\nu_x}$ & 1/2 & 2/3  & (2+m)/(2+2m) & 2/(d+2)
\\ \hline $\beta_h$, $\langle h \rangle \sim \delta \mu^{-\beta_h}$
& 0   & 1/3 & 1/(1+m)      & (2-d)/(d+2)\\ \hline $z$, $\xi \sim
t^{1/z}$ & 2   & 2    & 2 &  2
\\ \hline
\end{tabular}} 
\label{table1}

\newpage

\begin{center}
\LARGE{TABLE 2}
\end{center}
\vspace{5cm}

\centerline{
\begin{tabular}{|c||c|c||c|c||c||} \hline
\rule[3mm]{0mm}{1mm} {\it Exponent } 
& \multicolumn{2}{|c||}{\it $\lambda <0$} 
& \multicolumn{2}{|c||}{\it $\lambda >0$} 
& \rule[1mm]{0mm}{6mm} DP \cite{dp_exponents}
\\ \cline{2-5} 
&\rule[4mm]{0mm}{1mm} $h$ \cite{grasem} & $n$ \cite{genovese} & 
$h$ \cite{brazilian} & $n$ \cite{brazilian}& 
\\  
\hline \hline
$\theta_{h,n}$  & ---  & 1.1(1)  & 0.355(15) & 0.215(15) & 0.1595  \\ \hline
$\nu_x$   & ---  & 1.0(1)  & ---       & 0.97(5)   & 1.7338  \\ \hline
$\beta_{h,n}$   & 0.41 & 1.5(1)  & 0.52(2)   & 0.32(2)   & 0.2765  \\ \hline
$z$       & ---  & 1.52(3) & ---       & 1.55(5)   & 1.5807  \\ \hline
\end{tabular}}
\label{table2}

\newpage

\begin{center}
\LARGE{FIGURE CAPTIONS}
\end{center}
\vspace{0.9cm}

\noindent
Fig. 1: Separation $h({\bf x},t)$ of the  $A B$ interface
from a rigid wall.
\vspace{0.9cm}

\noindent
Fig. 2: Typical effective interfacial potentials for negative and
positive values of $b$. Below the transition temperature, the
potential well localizes the interface (dotted lines).
\vspace{0.9cm}

\noindent
Fig. 3: Left, pressure vs. temperature phase diagram of a pure
substance. $TP, CP$ and $T_w$ stand for triple point, critical
point, and wetting temperature; center, thickness of the wetting
layer as a function of the relevant parameters for the paths
indicated on the left and on the right; right, phase diagram in
the $b-\mu$ plane. Solid and dotted lines correspond to continuous
and first-order transitions, respectively (see text).
\vspace{0.9cm}

\noindent
Fig. 4: Phase diagrams for $\lambda <0$ (left) from numerical solutions 
of (\ref{bkpz}) in short-ranged potentials, and $\lambda >0$ (right) from 
discrete models.
The arrows denote different types of transitions explained in the text.
\vspace{0.9cm}

\noindent
Fig. 5: Snapshots of $d=2,1$ interfacial configurations within
the coexistence region, from numerical solutions of (\ref{bkpz}) in a
short-ranged potential.
\vspace{0.9cm}

\newpage 

\noindent
Fig. 6: Left, spatio-temporal evolution from a numerical solution
of (\ref{mn}) within the coexistence region. Depinned regions ($n<1$)
are colored in dark gray and pinned ones ($n>1$) in light gray.
The instantaneous interfacial configuration for the time marked
by the line is shown in figure (\ref{profiles}). Right,
single-site probability density function for different values of
$\mu$, also within the coexistence region. The stability edge
is $\mu^*(b=-4) \approx 1.3$.
\vspace{0.9cm}

\noindent
Fig. 7: Typical interface profile within the coexistence region 
and its associated particle representation. A long-ranged attraction 
between the edges of low-density segments tantamounts to an
a long-ranged attraction between the vertices forming the base of the 
triangle, which arguably induces an effective long-ranged attraction between 
the substrate and the interface.

\newpage

\center{\LARGE{FIGURE 1}}
\vspace{5cm}

\begin{center}
\includegraphics[angle=0,width=6cm]{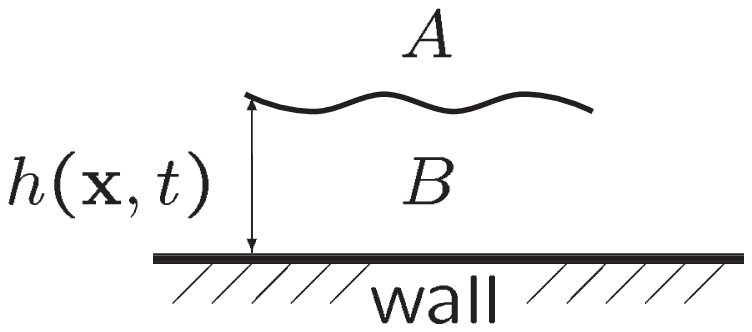}
\label{scheme}
\end{center}

\newpage

\center{\LARGE{FIGURE 2}}
\vspace{5cm}

\begin{center}
\includegraphics[angle=0,width=12cm]{rev_fig2.eps}
\label{potentials}
\end{center}

\newpage 

\center{\LARGE{FIGURE 3}}
\vspace{5cm}

\begin{center}
\includegraphics[angle=0,width=12cm]{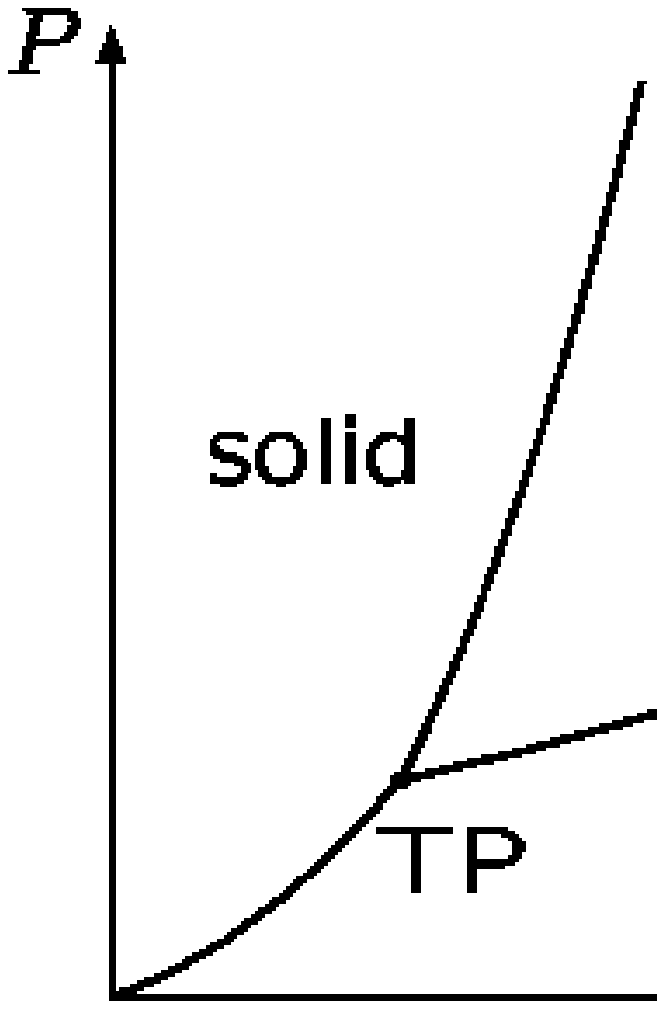}
\label{eq_pd}
\end{center}

\newpage 

\center{\LARGE{FIGURE 4}}
\vspace{5cm}

\begin{center}
\includegraphics[angle=0,width=12cm]{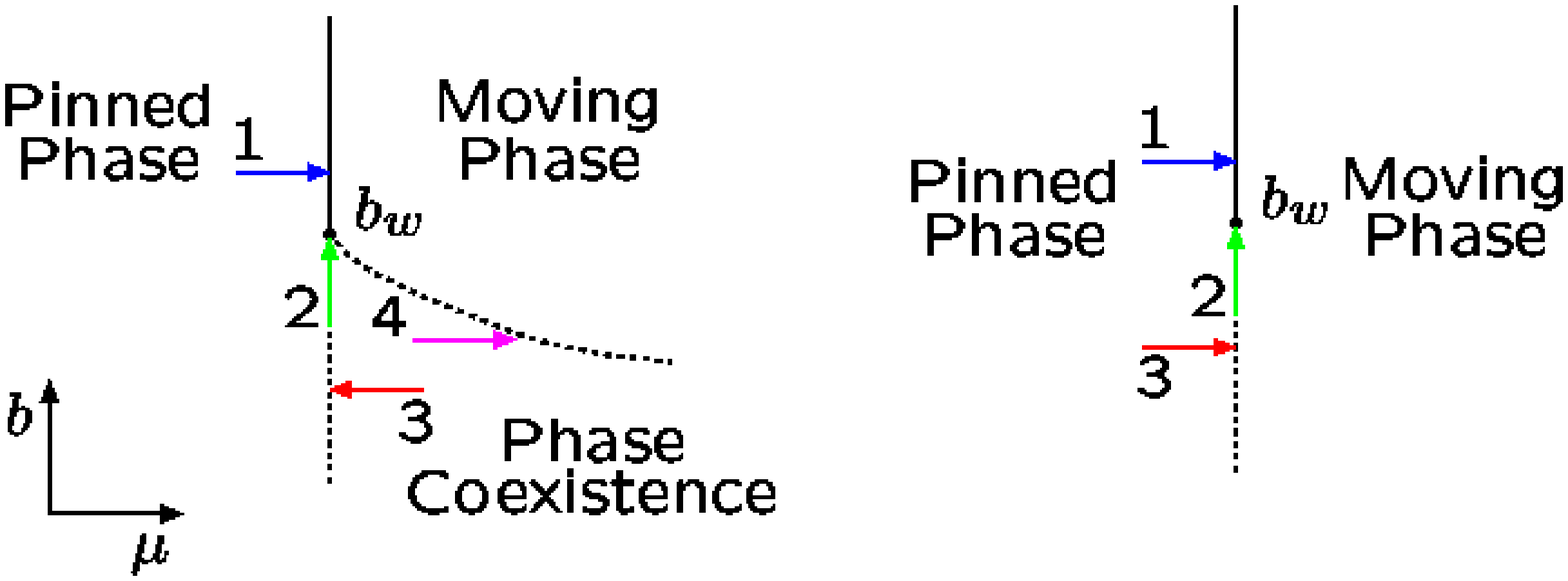}
\label{pd_sr}
\end{center}

\newpage

\center{\LARGE{FIGURE 5}}
\vspace{5cm}

\begin{center}
\includegraphics[angle=0,width=12cm]{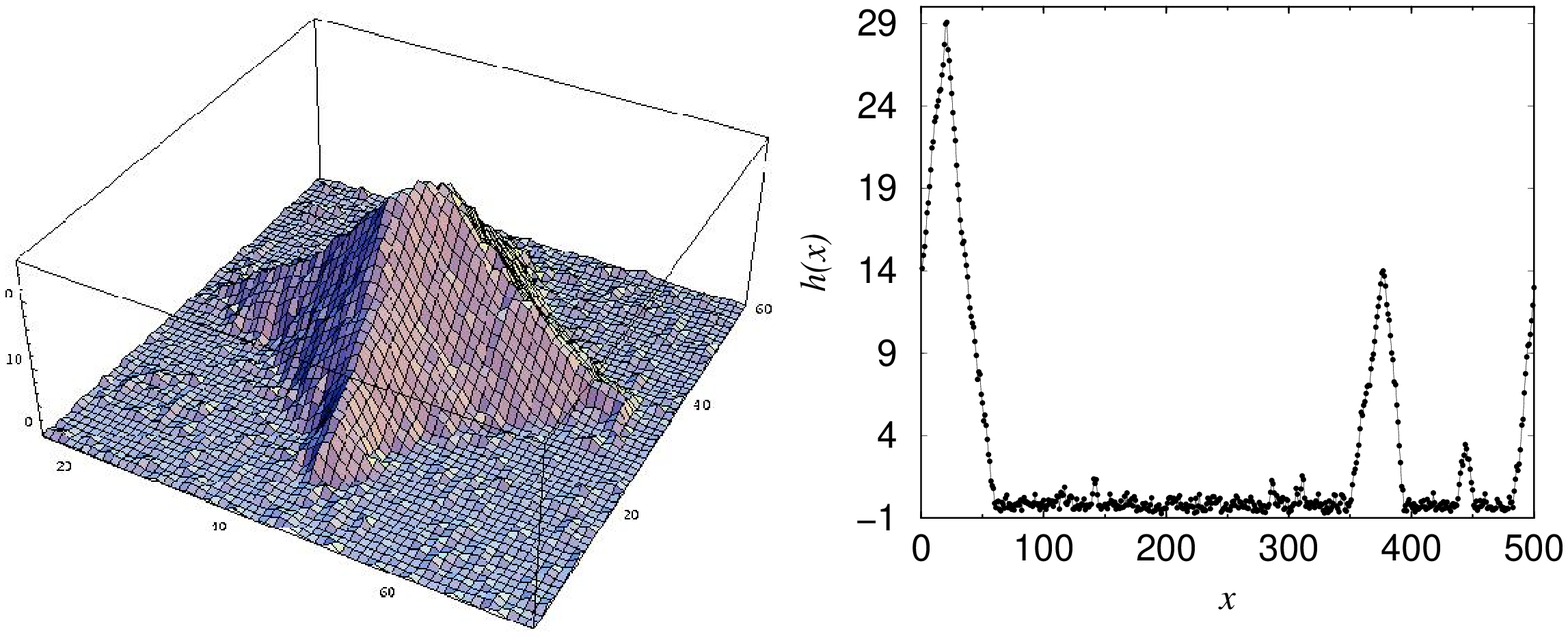}
\label{profiles}
\end{center}

\newpage

\center{\LARGE{FIGURE 6}}
\vspace{5cm}

\begin{center}
\includegraphics[angle=0,width=12cm]{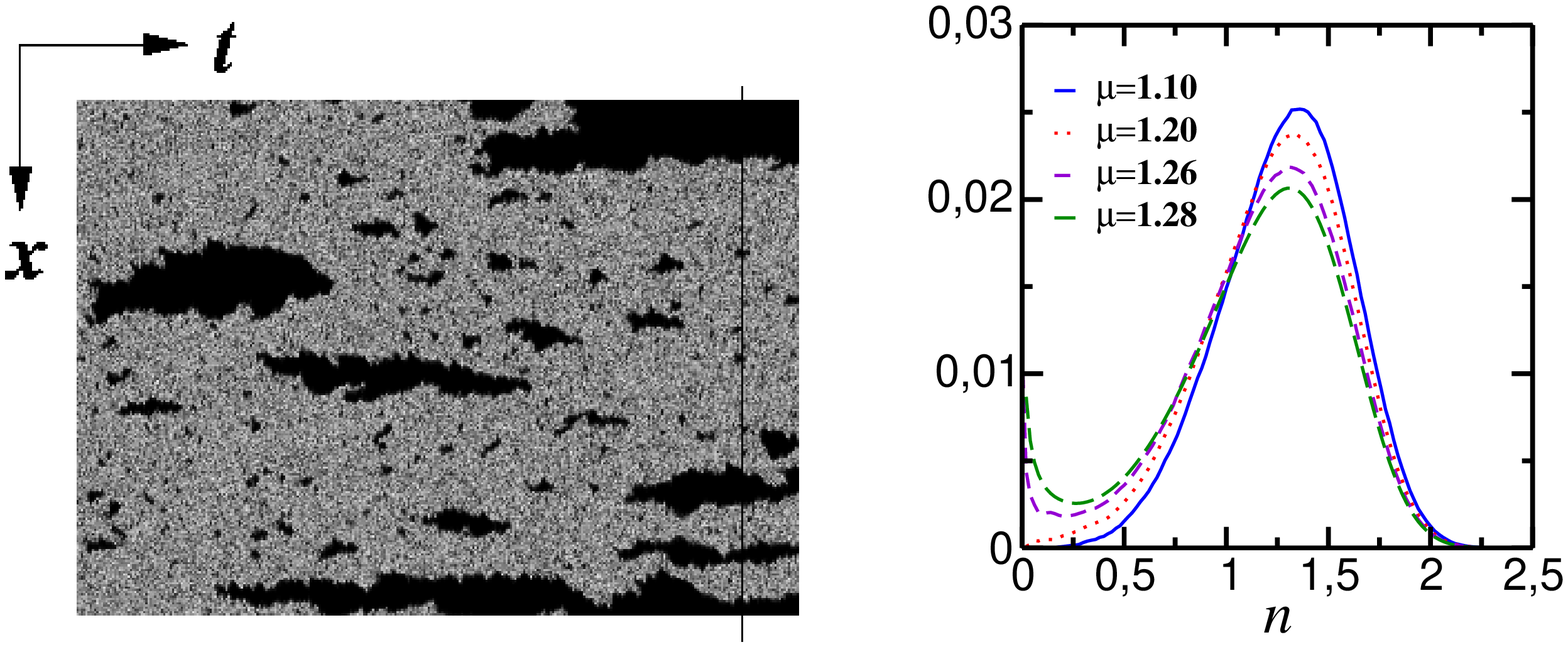}
\label{histo_conf}
\end{center}

\newpage

\center{\LARGE{FIGURE 7}}
\vspace{5cm}

\begin{center}
\includegraphics[angle=0,width=12cm]{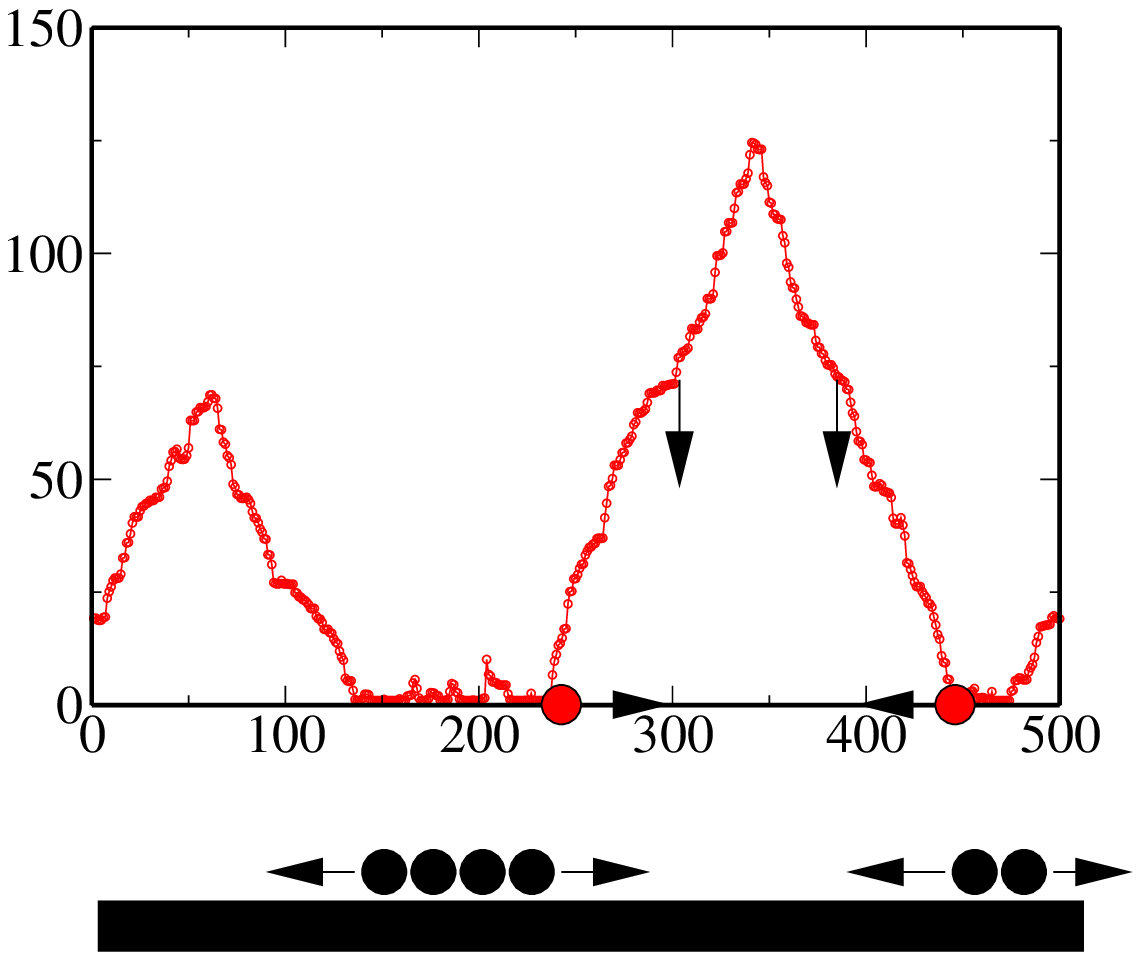}
\label{long_dp}
\end{center}

\end{document}